\documentclass[12pt,thmsa,a4paper]{article}

\usepackage{amsmath}
\usepackage{amssymb}
\usepackage{epsfig}
\usepackage[latin1]{inputenc}

\newcommand{\p}{\partial}
\newcommand{\n}{\nabla}

\newcommand{\Sss}{\left(1-\frac{2M}{r}\right)}
\newcommand{\Th}{T^{\theta}{}_{\theta}}
\newcommand{\E}[1]{\left< #1 \right>}

\newcommand{\vp}{\varphi}
\newcommand{\ve}{\varepsilon}

\newcommand{\be}{\begin{equation}}
\newcommand{\ee}{\end{equation}}
\newcommand{\ba}{\begin{eqnarray}}
\newcommand{\ea}{\end{eqnarray}}

\newcommand{\sgn}{\text{sgn}}
\newcommand{\D}{\Delta}
\newcommand{\sq}{\square}
\newcommand{\ga}{\alpha}
\newcommand{\gb}{\beta}
\newcommand{\gc}{\gamma}

\newcommand{\gl}{\lambda}

\begin{document}

\begin{titlepage}
\renewcommand{\thefootnote}{\fnsymbol{footnote}}

\hfill TUW--05--18 \\

\begin{center}

{\Large\bf IR Renormalisation of General Effective Actions and Hawking Flux
           in 2D Gravity Theories}\\
  \vspace{7ex}
  D. Hofmann,
  W. Kummer,
  \vspace{7ex}

  {\footnotesize Institut f\"ur
    Theoretische Physik \\ Technische Universit\"at Wien \\ Wiedner
    Hauptstr.  8--10, A-1040 Wien, Austria}
  \vspace{2ex}

\end{center}
\vspace{7ex}
\begin{abstract}
The infrared problem of the effective action in 2D is
discussed in the framework of the Covariant Perturbation
Theory. The divergences are regularised by a mass and the
leading term is evaluated up to the third order of perturbation theory.
A summation scheme is proposed which isolates
the divergences from the finite part of the series and
results in a single term. The latter turns out to be
equivalent to the coupling to a certain classical
external field. This suggests a renormalisation
by factorisation.
\end{abstract}


\vfill

\end{titlepage}

\section{Introduction}

In the last years two-dimensional models have widely been
used in the context of field quantisation of and upon curved
spacetimes \cite{gkv02}. By choosing an appropriate ``potential''
for the dilaton fields in the first order version of such
gravity models, essentially all physically interesting theories
can be covered, including interactions of gravity with matter, at least
as long as no explicit couplings of the matter (e.g. scalar fields)
to the scalar curvature is assumed.
In particular, Spherically Reduced Gravity (SRG) provides
a simple model to study four-dimensional Hawking radiation
by a two-dimensional dilaton action with just one dilaton
field \cite{hok05a,kuv99,ghk00a}. This is of special importance in connection
with the correct computation of the Hawking flux, which had been the subject
of some controversy \cite{boh97,kuv98}. It has been settled by
one of the present authors with D. V. Vassilevich \cite{kuv99},
who showed that the correct flux at infinity, resulting in the
Stefan-Boltzmann law as determined by the Hawking temperature, can be
obtained. On the other hand, a logarithmic divergence
(cf. also \cite{baf98,baf99}) of the flux
at the horizon (in global coordinates) seems to remain a difficulty,
at least for the fixed background of an ``eternal'' Black Hole.
 
The key problem has always been the computation of the
effective action, when the path integral of the
quantum field $S$ (we only consider a single scalar field)
is carried out. In previous work cited above \cite{kuv99}
this was treated by the heat-kernel method, but at one point applied
beyond its mathematically established range of applicability.

Another quite different approach uses the Covariant Perturbation Theory (CPTH)
of the functional determinant invented by Barvinsky and Vilkovisky
\cite{bav87,bav90a,bav90b}. We have shown recently that even for massless
scalar fields the expected result \cite{kuv99} for the
Hawking flux could be reproduced correctly using the CPTH in two dimensions \cite{hok05a}.
The drawback therein, however, has been the appearance of Infrared (IR)
divergences.

The IR problem in two dimensions has been
a matter of confusion since the invention of CPTH. The authors
of that method themselves claimed the non-analyticity and thus
non-applicability of their method in two dimensions, except for one
particular case (namely vanishing endomorphism, see below)
in which the major result, the trace anomaly, could
be derived by the local Seeley-DeWitt expansion as well
\cite{chf77,wal78}.
On the other hand, in other work \cite{guz00} the IR problem
also has been declared as non-existing\footnote{The origin of this
surprising result, which seems to contradict the analysis of Barvinsky and
Vilkovisky, might be that in their case orders in the (dimensionless!) dilaton
instead of orders in the curvature were considered.}.
Our viewpoint is that IR divergences \emph{exist} to all orders of CPTH,
\emph{however}, a procedure can be given to regularise and eventually
renormalise them by physical arguments.

In the present paper a regularisation procedure by a mass term $m^2$
for general effective actions in two dimensions is proposed.
We consider divergent terms $\ln m^2$ up
to the third order of CPTH and conjecture the possibility of a summation
of the series into a single term. Finally, we show how such
a term can be produced by an ambiguity of a source term coupled
to the scalar field $S$, representing some external classical field.

In Section 2 we shortly recall the main features of CPTH in two dimensions
and quote some specifications when applied to SRG.
The subsequent Section 3 is devoted to a formal analysis of
the divergences.
Section 4 contains the results of the mass-regularisation
of the effective action up to the third order of CPTH.
A possible summation of the divergences is conjectured.
In Section 5 we discuss the significance of the closed divergent
expression obtained in Section 4 and propose a renormalisation
by an external field.

The results are summarised in the Conclusions, where related
aspects and possible extensions are discussed as well.

The Appendices contain details of the calculations on which this
work is based. Appendix A presents the mass regularisation
for the second order of CPTH, Appendix B for the third order.
The meaning of formal terms of type $\ln\sq$ is discussed
in Appendix C.

\section{Effective Action in CPTH}

The effective action $W$ for a scalar field $S$
on a curved spacetime $L$ with metric $g$ is defined by
\be
e^{iW}={\cal N}\int{\cal D}\left({\sqrt[4]{-g}\tilde{S}}\right)
e^{-\frac{i}{2}\int_L\tilde{S}{\cal O}\tilde{S}\sqrt{-g}d^2x},
\label{gen-functional}
\ee
where the factor $\sqrt[4]{-g}$ in the path-integral measure
has been introduced to preserve general covariance \cite{fln88},
but can be eliminated right away by redefining $S:=\sqrt[4]{-g}\tilde{S}$.
The classical scalar field $S$ obeys the equation of motion
${\cal O}S=0$ with general d'Alembertian
\be
{\cal O}=\square+E.
\ee
$E$ (for endomorphism) refers to some potential coupled to $S$
which may contain one or several (dilaton) fields. It shall
not depend on $S$, excluding self-interaction. In SRG we have only
one dilaton field $\phi$, defined by $X=e^{-2\phi}$, where
$X$ may be gauged to simply represent the radius coordinate $r^2$
of the two-sphere in four dimensions. The corresponding endomorphism,
expressed by $\phi$, reads \cite{hok05a}
\be
E=\sq\phi-(\n\phi)^2.\label{E-SRG}
\ee
After introduction of Euclidean time
$\tau=i t$ and an Euclidean operator
${\cal O}_{\cal E}=-{\cal O}=\triangle+E_{\cal E}$ the effective action
(performing a Gaussian integration) can be written
as the derivative of the zeta-function\footnote{For simplicity we
shall write very often the Minkowski signature operator ${\cal O}$ in connection with the
zeta-function and the heat-kernel where there should be
$-{\cal O}_{\cal E}$.}
$\zeta_{{\cal O}_{\cal E}}[s]:=\text{tr}\left({\cal O}^{-s}\right)$
for the parameter $s$:
\be
W=\frac{i}{2}\ln\det(-{\cal O}_{\cal E})+\text{const.}
\approx\frac{i}{2}\text{tr}\ln(-{\cal O}_{\cal E})
=-\frac{i}{2}\frac{d}{ds}\zeta_{{\cal O}_{\cal E}}[s]\Bigm|_{s=0}
\ee
The zeta-function can be expressed by the heat-kernel $e^{{\cal O}_{\cal E}\tau}$
\be
\zeta_{{\cal O}_{\cal E}}[s]=\frac{1}{\Gamma(s)}\int_0^{\infty}\tau^{s-1}
\text{tr}\left(e^{{\cal O}_{\cal E}\tau}\right)d\tau.\label{zeta-function}
\ee
The trace of the heat-kernel in CPTH \cite{bav87} is expanded in orders of
curvature\footnote{In this context ``curvature'' means the scalar curvature $R$,
the endomorphism $E$ as well as some gauge curvature
(which in the present case is absent).}
\be
\text{tr}\left(e^{{\cal O}_{\cal E}\tau}\right)=\frac{1}{4\pi\tau}
\int_{\cal E}\left\{a_0+\tau a_1+\tau^2a_2+\dots\right\}\sqrt{g}d^2x_{\cal E},
\label{heat-kernel}
\ee
where the zeroth and first order coefficients are local and
agree with those of the Seeley-DeWitt expansion \cite{gil95}, namely
\be
a_0=1\,\,,\,\,a_1=\frac{R_{\cal E}}{6}+E_{\cal E}.\label{a_1,a_2}
\ee
All other coefficients are \emph{non-local}, i.e.
they contain integrations over the Green function on the
Euclidean spacetime $L_{\cal E}$, expressed by inverse powers
of $\sq$ (Green functions) and the presence of form factors like $f$
(cf. \cite{hok05a}, eq. 18):
\small
\begin{eqnarray}
a_2&=&R_{\cal E}\left[\frac{1}{16\tau\sq}+\frac{f(\tau\sq)}{32}+\frac{f(\tau\sq)-1}{8\tau\sq}
+\frac{3[f(\tau\sq)-1]}{8(\tau\sq)^2}\right]R_{\cal E}\nonumber\\
&&+E_{\cal E}\left[\frac{f(\tau\sq)}{6}+\frac{f(\tau\sq)-1}{2\tau\sq}\right]R_{\cal E}
+R_{\cal E}\frac{f(\tau\sq)}{12}E_{\cal E}+E_{\cal E}\frac{f(\tau\sq)}{2}E_{\cal E}
\label{a_2}\\
f(x)&=&\int_0^1e^{-a(1-a)x}da\label{f}
\end{eqnarray}
\normalsize
A series expansion of the effective action in orders of curvature
thus reads
\be
W=-\frac{i}{2}\frac{d}{ds}\left\{\frac{1}{\Gamma(s)}\int_0^{\infty}
\frac{d\tau}{4\pi\tau^{2-s}}\int_{\cal E}
\left\{a_0+\tau a_1+\tau^2a_2+\dots\right\}\sqrt{g}d^2x_{\cal E}\right\}\biggm|_{s=0}.
\label{eff-action}
\ee
The key feature of CPTH within the context of 2D dilaton gravity
is that the effective action
$W$ is computed directly from (\ref{eff-action}) (well-defined in the
Boulware state \cite{hok05a}), and not only certain functional
derivatives thereof in the conformal gauge, which are integrated later
on \cite{kuv99}. It is the (infinite) $\tau$-integral of (\ref{eff-action})
that makes the IR divergence explicit which in the previous approach
\cite{kuv99},\cite{kuv98} was not evident.

In order to be self-contained we quote the result of \cite{hok05a}
for the finite part (up to finite renormalisation effects from the UV/IR
divergences involving renormalisation constants $c_i$) of $W$ to
second order of CPTH (with Minkowski signature),
\be
W_{finite}=\frac{1}{96\pi}\int_{\cal M}\left[(R+12E)\frac{1}{\sq}R\right]\sqrt{-g}d^2x
\label{W-finite},
\ee
which for SRG, eq. (\ref{E-SRG}), reproduces the correct
asymptotic Hawking flux of ref. \cite{kuv99}.

\section{Divergences}

In this section we recall all divergences in W produced by CPTH
in two dimensions, including the UV divergence(s) of the lower order(s).
Generally one can say that a divergence at the lower limit
of the $\tau$-integration corresponds to a UV divergence,
while the upper limit may cause IR divergences.
In earlier work \cite{hok05a} we used a cut-off $T$ at large $\tau$
to control the IR behavior of the heat-kernel.
Here we introduce a mass-term instead, the advantage with respect to a
cut-off in $\tau$ being a clear separation between UV and IR divergences.

\subsection{UV Divergences}

From (\ref{eff-action}) we observe that
the zeroth order of CPTH, proportional to $a_0$, has a pole of first order
at $\tau=0$. In \cite{bav90a} this UV divergence had been
dismissed summarily by dimensional
regularisation arguments (see equation 3.17 in \cite{bav90a}).
In fact, a constant divergence $\propto1/\ve$,
where $\ve>0$ is a cut-off at $\tau=0$, can be interpreted
as the infinite contribution of the vacuum energy to the cosmological
constant:
\be
W_{UV}=W_0=\frac{1}{8\pi}\int_{\cal M}\frac{1}{\ve}\sqrt{-g}d^2x
\ee
Its infinite contribution to the Hawking flux in SRG could be
renormalised by substracting the flat spacetime value \cite{hok05a}.

Apart from the one related to $a_0$ there are no further
UV divergences in two dimensions.

\subsection{IR Divergences}

Starting with the first one all orders in CPTH produce IR divergences.
It is obvious that the divergence of the first order is
logarithmic and simply proportional to $a_1$.
The higher order coefficients include exponential form
factors like e.g. (\ref{f}), leading to logarithmic divergences as well.

It should be noted that for the particular case of $E=0$ (providing conformal
invariance of the two-dimensional action),
the effective action becomes IR finite even in two dimensions \cite{bgv94}!

The effective action shows a well-known ambiguity due to
translation invariance of the path integral which can be expressed
in the form
\be
W=-\frac{i}{2}\frac{d}{ds}\zeta_{{\cal O}_{\cal E}}[s]\Bigm|_{s=0}
-\frac{i}{2}\zeta_{{\cal O}_{\cal E}}[0]\ln\tilde{\mu}^2.
\label{amb}
\ee
Its origin is the possibility\footnote{We emphasize already here
that this possibility disappears in the presence of some external
source (cf. the last paragraph of Section 5).} to redefine $\cal O$ by
a multiplication with $\tilde{\mu}^{-2}$. This
can be shifted into the scalar field $S$ and simply
leads to a multiplication of all positive eigenvalues of the elliptic
operator $-{\cal O}_{\cal E}$
by this factor and thus to a contribution to the effective action
because $\zeta[s]=\text{tr}(\lambda^{-s})\to\text{tr}[(\lambda/\tilde{\mu}^2)^{-s}]$
\cite{gkv02}. In the present case we have
$\zeta_{{\cal O}_{\cal E}}[0]=\frac{1}{4\pi}\int_{L_{\cal E}}a_1\sqrt{g}d^2x_{\cal E}$
and hence by (\ref{a_1,a_2}) an ambiguity
\be
W_{amb}=-\frac{1}{96\pi}\int_{\cal M}\left[(2R+12E)\ln\tilde{\mu}^2\right]\sqrt{-g}d^2x.
\label{ambiguity}
\ee
Therefore, at first sight the IR divergence (\ref{first-order})
of the first order CPTH could
be renormalised by simply adjusting the constant $\tilde{\mu}^2$.
However, the latter could be related to the UV renormalisation
in higher-dimensional theories (as usual in two dimensions UV and
IR divergences may combine). Anyway, an ambiguity of the type
(\ref{ambiguity}) is not sufficient to renormalise \emph{all}
orders of CPTH.

\section{IR Regularisation}

We regularise the effective action by introducing a mass-term
in the d'Alembertian: ${\cal O}\to{\cal O}+m^2$. This mass-term
can be pulled out from the trace of the heat-kernel
\be
\text{tr}\left(e^{-\left[{\cal O}+m^2\right]\tau}\right)
=e^{-m^2\tau}\text{tr}\left(e^{{\cal O}_{\cal E}\tau}\right),
\label{mass-term}
\ee
and thus regularises the zeta-function
(\ref{zeta-function}) at the upper limit of $\tau$.

The leading power in the radius $r$ which determines the asymptotic flux
for SRG from (\ref{W-finite}) from the second order in CPTH has been
found to be independent of the IR problem. In order to try
to understand this we thus have to go beyond that.
In the following we compute the IR divergences up to the third
order of CPTH. The final expressions are presented
in Minkowski signature spacetime. Thereby we pick up a factor $i$
from the Euclidean volume element $d^2x_{\cal E}=i\cdot d^2x$
which is multiplied by the factor $-i$ in (\ref{eff-action})
resulting in \emph{no} overall sign change. The Euclidean expressions
in the zeta-function $\zeta[s]$ like metrics, scalar curvature
etc. produce a minus sign when switching to Minkowski signature spacetime.

\subsection{First Order of CPTH}

In the first order term $\propto a_1$ a simple substitution of the
integration variable $\tau\to m^2\tau$ is sufficient
\be
\frac{d}{ds}\left\{\frac{1}{\Gamma(s)}\int_0^{\infty}
\tau^{s-1}e^{-m^2\tau}d\tau\right\}
=\frac{d}{ds}\left\{(m^2)^{-s}\right\}=-(m^2)^{-s}\ln m^2,
\ee
yielding at $s\to0$ the expected $\ln m^2$ behavior.
Going back to Minkowski signature spacetime the resulting contribution $W_1$
to the effective action reads
\be
W_1=\frac{1}{96\pi}\int_{\cal M}\left[(2R+12E)\ln m^2\right]\sqrt{-g}d^2x.
\label{first-order}
\ee
Replacing all Euclidean by Minkowski signature variables
the curvatures acquire a minus sign by this transition:
$R=-R_{\cal E},E=-E_{\cal E}$. Therefore, the choice $\mu^2=m^2$
in (\ref{ambiguity}) would be sufficient to formally
renormalise the first order divergence to zero. However,
this would contribute to the inherent mixing of UV and IR renormalisation
alluded to already above which we want to avoid.

Of course, $R\sqrt{-g}$ is a total divergence in two
dimensions and we therefore may omit its contribution to the
effective action.

\subsection{Second Order of CPTH}

The second order (\ref{a_2}) of CPTH consists of non-local terms only
that are produced by the form-factor $f$, eq. (\ref{f}). As announced
we write $\square$ instead of $-\triangle$ for simplicity,
although we should still treat quantities to be Euclidean at
intermediate steps. There are five types of heat-kernel
integrals contributing to the second order in (\ref{a_2}),
the most non-trivial one being (see Appendix A,
eqs. (\ref{I_f-exact}),(\ref{F}))
\begin{multline}
I_f=\frac{d}{ds}\left\{\frac{1}{\Gamma(s)}\int_0^{\infty}
\tau^sf(\tau\sq)e^{-m^2\tau}d\tau\right\}\biggm|_{s=0}
=\frac{1}{m^2}F\left(\frac{\sq}{m^2}\right)\\
=\frac{z\cdot F(z)}{\sq}=\frac{2}{\sq}\frac{\text{Arcosh}
\left(\frac{z}{2}+1\right)}{\sqrt{1+\frac{4}{z}}},
\label{integral-1}
\end{multline}
where $z=\sq/m^2$. $F(z)$ is a regular function for all $z\in[0,\infty[$ and possesses a
series expansion around $z=0$. This allows for an expansion of
$F(\sq/m^2)$ in terms of a complete set of eigenfunctions\footnote{For
SRG the dimensionless variable $y=\frac{r}{2M}-1$ is related to the
radius measured from the horizon $2M$, $\gl\ge0$ is a properly defined
dimensionless eigenvalue (cf. Appendix C).}
$\vp_{\gl}(y)$ of $\sq$ which proves useful when examining the
action of (\ref{integral-1}) on some function, say $E$ as in (\ref{a_2}):
\begin{multline}
I_f\,E=\frac{z\cdot F(z)}{\sq}\int_0^{\infty}\delta(y-y')E(y')dy'\\
=\frac{1}{\sq}\int_0^{\infty}d\gl\left(\frac{\sq}{m^2}\right)\cdot F\left(\frac{\sq}{m^2}\right)
\vp_{\gl}(y)\int_0^{\infty}\vp_{\gl}(y')\cdot E(y')dy'\\
=\frac{4M^2}{\sq}\int_0^{\infty}d\gl\frac{2\text{Arcosh}\left(\frac{\frac{\gl^2}{4M^2m^2}+2}{2}\right)}
{\sqrt{1+\frac{16M^2m^2}{\gl^2}}}\,\,\vp_{\gl}(y)\int_0^{\infty}
\vp_{\gl}(y')\cdot E(y')dy'\label{I_f E}
\end{multline}
Here and in the following it is convenient to keep $\sq^{-1}$ outside
the $\gl$-integral. This seems natural since all (IR-regular)
second order terms contain one Green function making them non-local.
Its action on the whole expression should be determined by different means,
e.g. by letting it act to the left (cf. (\ref{heat-kernel}) with (\ref{a_2})).

In a next step we take the limit $m\to0$ at fixed, finite $\gl$:
\be
\frac{\gl}{m^2}F\left(\frac{\gl}{m^2}\right)\stackrel{m\to0}{\to}
-2\ln\left(\frac{m^2}{\gl}\right)+O(m^2).\label{limit}
\ee
Whether this really spearates the IR divergence $\propto\ln m^2$
clearly depends on the behavior of the integrand in (\ref{I_f E})
at $\gl\to0$ for which the dependance on $\gl$ of the eigenfunctions
is crucial. Another way to check the justification of this
separation for all values of $\gl$ consists in assuming (\ref{limit})
and to verify that the finite remnant, essentially given by
$\sq^{-1}\int_0^{\infty}d\gl\ln\gl\cdot\vp_{\gl}(y)\int_y\vp_{\gl}(y')E(y')dy'$,
is well-defined. An analysis (Appendix C) of the eigenfunctions and of the
double integral in (\ref{I_f E}) suggests that the latter
converges for all values of $y$ and in the case of SRG falls off asymptotically
at least as $y^{-1}$. Therefore, the limit (\ref{limit})
can be justified indeed in (\ref{integral-1}) even including $\gl\to0$,
yielding the formal limit:
\be
I_f\stackrel{m\to0}{\to}-2\frac{\ln\left(\frac{m^2}{\sq}\right)}{\sq}+O(m^2)
\label{I_f limit}
\ee
In \cite{hok05a} we already showed that only the last term
in (\ref{a_2}), which is of the type $E_{\cal E}\,I_f\,E_{\cal E}$, leads to the IR
divergence of the second order CPTH. Although $I_f$ also appears
in other terms in (\ref{a_2}) which are of the type
$R_{\cal E}^2\,,\,R_{\cal E}\cdot E_{\cal E}$ those contributions cancel.
In comparison with the cut-off regularisation $\tau\le T$ of \cite{hok05a}
the result (\ref{I_f limit}) is shifted by some finite constant
$\ln m^2=-\ln T-\gc_E$ where $\gc_E\approx0.57721$ is the Euler constant.
To check, whether the regular part of the second order CPTH
remains unchanged, also the integral
(see Appendix A, eq. (\ref{G}))
\be
I_{f-1}=\frac{d}{ds}\left\{\frac{1}{\Gamma(s)}\int_0^{\infty}\tau^{s-1}
\frac{\int_0^1e^{-a(1-a)\sq\tau}da-1}{\sq}e^{-m^2\tau}
d\tau\right\}\biggm|_{s=0}
=-\frac{z\cdot G(z)}{\sq}
\label{integral-2}
\ee
is needed. Again the formal limit $m\to0$ is justified:
\be
I_{f-1}\stackrel{m\to0}{\to}\frac{\ln\left(\frac{m^2}{\sq}\right)+2}{\sq}+O(m^2)
\ee
Comparing with $\tau$-regularisation in \cite{hok05a}, eq. (\ref{integral-2}) is
shifted by the same amount as in (\ref{I_f limit}), and therefore
(one piece\footnote{There is another \emph{regular} non-local contribution
$\propto R^2_{\cal E}$ to this order of CPTH \cite{bav90a}
(the first line in (\ref{a_2})) which is not considered here.} of) the
regular part is in perfect agreement with \cite{hok05a} (the terms
$\propto E_{\cal E}\cdot R_{\cal E}$ in (\ref{a_2})):
\be
\frac{d}{ds}\left\{\frac{1}{\Gamma(s)}\int_0^{\infty}\tau^{s-1}
d\tau\right\}E_{\cal E}\left[\frac{f(\tau\sq)}{4}
+\frac{f(\tau\sq)-1}{2\tau\sq}\right]R_{\cal E}
=E_{\cal E}\frac{1}{\sq}R_{\cal E}.\label{regular-part}
\ee
This proves the concordance of the mass-regularisation
with the cut-off one in the regular sector,
yielding the IR divergent contribution
\be
W_2=\frac{1}{96\pi}\int_{\cal M}\left[-
12\cdot E\frac{\ln\left(\frac{m^2}{\sq}\right)}{\sq}E\right]
\sqrt{-g}d^2x\label{second-order}
\ee
to the effective action.

\subsection{Third Order of CPTH and Summation}

To third order of CPTH technical complications increase dramatically.
Therefore, we make the assumption that the IR divergence is contained
in the ``pure'' endomorphism term which is $\propto E^3$.
This is supported by the fact that
\begin{itemize}
\item the IR divergences were ``purely endomorphism''
to the first two orders;
\item the effective action becomes IR finite in the case
$E=0$ \cite{bgv94}.
\end{itemize}
In Appendix B we show that (under this assumption) the unique, logarithmically
divergent part of the endomorphism term is given by
\be
W_3=\frac{1}{96\pi}\int_{\cal M}\left[12\cdot\ln m^2\cdot
E\frac{1}{\sq}E\frac{1}{\sq}E\right]\sqrt{-g}d^2x.\label{third-order}
\ee
Here we have disregarded terms formally written as $\ln\sq$ as
we are mainly interested in the IR structure.
The proper interpretation of such terms, being IR regular,
is discussed in Appendix C.

Inspection of the IR divergent terms (\ref{first-order}), (\ref{second-order}),
and (\ref{third-order}) reveals a nice pattern, suggesting a similar structure
of the higher order ones in an infinite series, which even lends
itself to a formal summation.
It proves useful to introduce some renormalisation parameter $\mu^2$
by replacing $\ln m^2=\ln(m^2/\mu^2)-\ln\mu^2$ and
$\ln(m^2/\sq)=\ln(m^2/\mu^2)-\ln(\mu^2/\sq)$, respectively.
Adding (\ref{first-order}), (\ref{second-order}), and (\ref{third-order})
the IR-divergence of the effective action (up to the third order
of CPTH) can be given in the form
\be
W_{IR}=\frac{1}{8\pi}\int_{\cal M}\left[
\left(E-E\frac{1}{\sq}E+E\frac{1}{\sq}E\frac{1}{\sq}E\right)
\cdot\ln\frac{m^2}{\mu^2}\right]\sqrt{-g}d^2x.
\ee
It is striking that these terms can be reproduced by the formal
series expansion
\begin{multline}
\sq\,\frac{1}{\cal O}\,E=\sq\,\frac{1}{\sq+E}\,E
=\sq\,\frac{1}{1+\frac{1}{\sq}E}\,\frac{1}{\sq}E\\
=\sq\left(1-\frac{1}{\sq}E+\frac{1}{\sq}E\frac{1}{\sq}E+\dots\right)\frac{1}{\sq}E\\
=E-E\frac{1}{\sq}E+E\frac{1}{\sq}E\frac{1}{\sq}E+\dots\,\,.\label{series}
\end{multline}
This suggests that the total IR divergence of the effective
action can be represented formally as
\be
W_{IR}=-\frac{1}{8\pi}\int_{\cal M}\left\{\sq\,\frac{1}{\cal O}\,E\right\}
\cdot\ln\xi\,\,\sqrt{-g}d^2x,\label{IR-divergence}
\ee
where $\xi=\mu^2/m^2\to+\infty$ in the limit $m^2\to0$.
Expression (\ref{IR-divergence}) indeed produces divergent contributions
to expectation values as for instance the ``dilaton anomaly''
in SRG \cite{kuv99}:
\be
\E{\Th}_2\propto\frac{\delta W_{IR}}{\delta\phi}
=\int_{\cal M}\frac{\delta W_{IR}}{\delta E}\frac{\delta E}{\delta\phi}
\sqrt{-g}d^2x
=\frac{1}{8\pi}\int_{\cal M}\frac{\delta E}{\delta\phi}
\frac{\ln\xi}{(1+\rho)^2}\sqrt{-g}d^2x
\ee
Here we have used the SRG conformal gauge\footnote{Conformal gauge
is defined by $g_{\ga\gb}=e^{2\rho}\eta_{\ga\gb}$ generally and
$\rho=\ln\Sss/2$ in the particular case of SRG. The implicit dependance
on the tortoise coordinate $r(r_{\ast})$ is not relevant here.}
representation $E=2M/r^3=\sq\rho$
(implying $\phi=-\ln r$ in (\ref{E-SRG})) \cite{hok05a}.
However, as it should be, the IR divergence (\ref{IR-divergence}) does \emph{not}
contribute to the trace anomaly\footnote{In this context note that
the representation $E=\sq\rho$ of the endomorphism can be
used only \emph{after} variation of the effective action
(before that we must use (\ref{E-SRG})).
Therefore, (\ref{IR-divergence}) does not contain the conformal factor $\rho$
explicitly and hence is conformally invariant.}.

Of course, a mathematically stringent discussion of the convergence for
the series expansion (\ref{series}) is impossible.
Nevertheless, we add here some heuristic argument.
A necessary formal condition is $|\sq^{-1}E|<1$. In the case of SRG
we have $\sq^{-1}E=\rho=\ln\Sss/2$, and therefore
this condition is fulfilled for
$r>1.156\cdot r_h$ where $r_h=2M$ is the radius of the
event horizon. Thus the series may converge a short distance outside
the horizon, excluding thereby, however, the most interesting region.
On the other hand, it could be speculated that the logarithmic
divergence at the horizon \cite{kuv98,baf98,baf99} when the leading
flux term is extrapolated back may be related to
(or even compensated by) a divergence of the series (\ref{series}).

\section{Renormalisation}

We have shown, that the formal expression (\ref{IR-divergence})
correctly reproduces the IR divergences of the CPTH
up to the second order and at least partly to the third order.
The next task is whether one can find physically
reasonable counterterms in order to eliminate (\ref{IR-divergence}).

To this account we introduce a source term $jS$ beside the scalar
action in the path integral (\ref{gen-functional}):
\be
L[g,S]=\int_{\cal M}\left[jS-\frac{1}{2}S{\cal O}S\right]\sqrt{-g}d^2x
:=-\frac{1}{2}\int_{\cal M}\left[\hat{S}{\cal O}\hat{S}
+j{\cal O}^{-1}j\right]\sqrt{-g}d^2x.\label{class.action}
\ee
The current $j$ has been shifted into the redefined scalar field
$\hat{S}$ by translation invariance of the path integral.
Now, by rewriting the inverse operator as
${\cal O}^{-1}=\sq^{-1}\sq{\cal O}^{-1}\sq\sq^{-1}$, where
$\sq{\cal O}^{-1}\sq$ can be expanded in a series similar to (\ref{series}),
one obtains for the source term $W_j$ of the effective action
\small
\begin{multline}
W_j[g,j]=-\frac{1}{2}\int_{\cal M}j\frac{1}{\sq}\left(\sq\frac{1}{\cal O}\sq\right)
\frac{1}{\sq}j\sqrt{-g}d^2x\\
=-\frac{1}{2}\int_x\int_y\int_zj(x)G(x,y)
\left(\sq-E+E\frac{1}{\sq}E-\dots\right)_yG(y,z)j(z)\\
=-\frac{1}{2}\int_{\cal M}j\frac{1}{\sq}j\sqrt{-g}d^2x\\
+\frac{1}{2}\int_x\int_y\int_z\left(E-E\frac{1}{\sq}E+\dots\right)_y
G(y,x)j(x)G(y,z)j(z)\\
=-\frac{1}{2}\int_{\cal M}j\frac{1}{\sq}j\sqrt{-g}d^2x
+\frac{1}{2}\int_{\cal M}\left(E-E\frac{1}{\sq}E+\dots\right)
\left(\frac{1}{\sq}j\right)\left(\frac{1}{\sq}j\right)\sqrt{-g}d^2y.
\label{W_j}
\end{multline}
\normalsize
For clarity in the intermediate step of (\ref{W_j}) we wrote the Green
function $G(x,y)$, as defined by $\sq G(x,y)=-\delta^2(x-y)$,
instead of $-\sq^{-1}$. We furthermore
assumed it to be symmetric in its arguments\footnote{This restricts
our argument to the Feynman Green function which, anyway,
is naturally preferred due to the Euclidean analysis
in the derivation of the effective action. Here we differ from
Barvinsky and Vilkovisky who argue that the retarded Green function
should be inserted ``by hand'' \cite{bav87}. In the context of SRG this
step could not be confirmed \cite{hok05a}.}.
We now redefine the source $j$ by adding a term $-\sq\chi_0$.
Then, up to terms that vanish in the limit $j\to0$, a contribution quadratic
in $\chi_0$ survives:
\be
W_j[g,j]=\frac{1}{2}\int_{\cal M}\left[\left(E-E\frac{1}{\sq}E+\dots\right)\chi_0^2
-\chi_0\sq\chi_0 \right]\,\,
\sqrt{-g}d^2x+O(j),
\label{ambiguity2}
\ee
where the terms in the brackets are precisely of the form (\ref{series})
of the IR divergence (\ref{IR-divergence}). Thus the latter can be removed to
all orders of CPTH by the choice
\be
\chi_0^2:=\frac{\ln\xi}{4\pi}=\frac{\ln\left(\frac{\mu^2}{m^2}\right)}{4\pi}>0.
\ee
We emphasize that even the sign is consistent as $\chi_0^2$ clearly must be positive.
For such a zero-mode $\chi_0$ the additional term $\propto\chi_0\sq\chi_0$
in (\ref{ambiguity2}) vanishes.

As for constant $\chi_0$ we have trivially $\sq\chi_0=0$ at first sight
our procedure seems to be very strange. Indeed,
reconsidering the classical action (\ref{class.action})
with source term $jS$ \emph{before redefining the scalar field $S$}
a shift in the source $j\to j-\sq\chi_0$ leaves it invariant
if $\chi_0$ is a constant (or a zero-mode).
However, this shift affects the effective action containing
combinations $\sq^{-1}j$, i.e. a non-local effect occurs.
Therefore, the contribution (\ref{ambiguity2}) to the effective action
must be considered some quantum ambiguity similar to the
well-known one described in (\ref{amb}) and (\ref{ambiguity}) above.
There we saw that the latter could be used to renormalise (\ref{first-order}),
the first order of CPTH, only.
In this respect the present ambiguity appears to be an extension
of that to all orders of CPTH, whereby the origin
again could be found in the translation invariance
of the path integral. 
It is, however, powerful enough to remove \emph{all} IR
divergences of the theory (assuming that our conjecture upon
the higher order terms is valid). Thereby it introduces
the renormalisation constant $\mu^2$ into the remaining
finite part of the effective action which should be determined
by the value of the Hawking flux.

A crucial difference between the former (\ref{ambiguity}) and the new
ambiguity (\ref{ambiguity2}) is the relation to a source
term in the case of the latter. In the presence of this
source term the former ``symmetry'' of the path integral
under multiplication of $\cal O$ by some renormalisation constant,
which led to the original ambiguity (\ref{ambiguity}), is destroyed.
Therefore it seems that one has to choose between
two kinds of ambiguities, whereby only the one exhibited
here allows a complete IR renormalisation to all orders of CPTH
which works by factorising out the IR-terms.

\section{Conclusions}

The aim of this paper was to shed new light on the infrared problem
inherent in general scalar effective actions in two
dimensions. To establish the effective action we used
the Covariant Perturbation Theory of Barvinsky and Vilkovisky \cite{bav87},
based upon an expansion in orders of the curvature.
Curvature in this context not only refers to the scalar
curvature $R$ associated with the d'Alembertian $\sq$ but
also to some potential $E$ called endomorphism ``acting'' on
the scalar field $S$ and thus forming a general d'Alembertian
${\cal O}=\sq+E$.

We further added a mass-term to control
the infrared divergences of the effective action (\ref{mass-term}).
The calculation of these divergences has been performed in detail for
the second and, with simplifying assumptions, also to third order of
Covariant Perturbation Theory. In comparison with a previous
approach \cite{hok05a} (where a cut-off of the eigentime $\tau$ in
the heat-kernel formalism was used as a regulator) the results for the
integrals were the same up to a finite constant. The latter is
irrelevant as it does not enter the regular part of the effective action (\ref{regular-part}).
The key problem has been the separation of the infrared divergences
proportional to $\ln m^2$ which, starting with the second order of
Covariant Perturbation Theory, are coupled to formal terms $\ln\sq$.
Going back to the origin of the latter and using an expansion
in terms of eigenfunctions of (the radial part of) the d'Alembertian
we could show that the separation of the IR divergences
produced finite remnants of the same type, including some
renormalisation constant $\mu^2$.

The infrared divergences of the first
three orders of Covariant Perturbation Theory were shown to
fit into the pattern of a formal series (\ref{series})
that could be summed into a single term (\ref{IR-divergence}),
containing the inverse of the general d'Alembertian ${\cal O}$.
At this point we conjectured the extension of this
identity to higher orders.

The introduction of a source term $jS$ (\ref{class.action})
in the path integral revealed the existence of an ambiguity
(\ref{ambiguity2}) of the effective action due to
its non-locality. A shift in the source $j\to j-\sq\chi_0$
leaves the classical action invariant but produces
non-vanishing contributions to the effective action if $\chi_0$
is a zero-mode of the d'Alembertian. The resulting contribution
exhibits the same structure as the infrared divergence (\ref{IR-divergence})
but carries the opposite sign and thus proves a good candidate
for renormalisation. Indeed, by a choice $\chi_0^2=\ln\xi/4\pi$
where $\xi=\mu^2/m^2>0$, the total infrared divergence of
the effective action can be renormalised, thereby leaving finite
terms containing some constant $\mu^2$. The latter appear
in new types of terms in the effective action which had not been
considered previously ($\propto \ln\sq/\mu^2$), and should be fixed by physical
observables like the Hawking flux in the case of spherically
reduced gravity.

A particular feature of these renormalisation terms is
that they are \emph{not} conformally invariant and thus
lead to contributions to the trace anomaly.
A heuristic argument suggests that they contribute significantly
only in the region where the convergence of the series
(\ref{series}) breaks down (i.e. close to the horizon).
There they may even lead to a logarithmic divergence, similar
to the one found in the ``dilaton anomaly'' \cite{kuv99}.
This behavior might hint at an analogous problem of
convergence in the regular sector of the Covariant Perturbation
Theory, which, however, is more difficult to investigate
since the coefficients show no comparable, simple pattern.
On the other hand it might even be that the logarithmic divergences
produced by the renormalisation terms compensate the one
of the dilaton anomaly.

For vanishing endomorphism $E$ the infrared divergence (\ref{IR-divergence}) becomes
a trivial surface term and hence vanishes. This
is in agreement with the observation that the effective
action should be finite in two dimensions in this particular
case \cite{bgv94}.

It is a peculiarity of our approach that the shift in the source
$j$ produces a term linear in $j$ and thus a non-vanishing
expectation value $\E{S}$. By construction it is a solution
of the d'Alembertian $\cal O$ because ${\cal O}\E{S}=\sq\chi_0=0$ and
approaches asymptotically the zero-mode $\E{S}\stackrel{r\to\infty}{\to}\chi_0$.

The present calculations clearly cannot prove the assumptions
made about the summation of the IR divergences to all orders
as they are based upon arguments only involving terms up to the third order.
However, the cancellation of divergences up to the second and (at least partly)
to the third order by means of our approach is manifest and thus
strongly supports the results on the Hawking flux obtained
by this method \cite{hok05a}.

We found that the IR divergent terms, as well as their compensation,
can be collected in a factor. Such factorisations
of IR terms in the limit of small masses are a phenomenon
which is known for a long time in connection with the emission of soft photons
\cite{bln37,yfs61}. It is amusing, that a similar phenomenon seems
to occur here and may suggest an analogous renormalisation prescription.
\\
\\
\textbf{\large Acknowledgement}
\\
\\
The auhors have profited from discussions with H. Balasin.
This work was supported by the Austrian Science Foundation (FWF)
Project P17938-N08 TPH.

\begin{appendix}

\section{Heat-Kernel Integrals to the Second Order of CPTH
- Mass Term Regularisation}

The exponential function in $f(\tau\sq)$ in (\ref{integral-1})
(which in the following we call $I_f$) can be expanded formally
in a series
\be
I_f=\sum_{n=0}^{\infty}(-1)^n\frac{c_n}{n\text{!}}\sq^n\frac{d}{ds}
\left\{\frac{1}{\Gamma(s)}\int_0^{\infty}
\tau^{s+n}e^{-m^2\tau}d\tau\right\}\biggm|_{s=0},
\ee
where $c_n=\int_0^1[a(1-a)]^nda=(1/4)^n\int_0^1(1-u^2)^ndu$.
In the integral
\be
C_n(k):=\int_0^k(k^2-u^2)^ndu=k^{2n+1}C_n(1)
\ee
differentiation for $k$ gives
\be
\frac{d C_n(k)}{d k}=\int_0^kn(k^2-u^2)^{n-1}2k\,du,
=(2n+1)k^{2n}C_n(1)
\ee
and for $k=1$ the recursion formula
\be
(2n+1)C_n(1)=2n\cdot C_{n-1}(1).
\ee
Since $c_n=(1/4)^nC_n(1)$ this becomes a similar relation for the $c_n$:
\be
c_{n+1}=\frac{n+1}{4n+6}\cdot c_n\label{recursion}
\ee
Using $\int_0^{\infty}e^{-m^2\tau}\tau^{s+n}d\tau=\frac{\Gamma(s+n+1)}{m^{2(s+n+1)}}$
we can perform the $\tau$-integration of $I_f$:
\be
I_f=\frac{1}{m^2}\sum_{n=0}^{\infty}(-1)^nc_nz^n
:=\frac{z\cdot F(z)}{\sq}\label{I_f-exact},
\ee
where we have defined the function $F(z)$ and $z:=\sq/m^2$.
In order to obtain a closed expression for $F$ the recursion relation
(\ref{recursion}) can be used to establish a differential equation
\be
F'(4z+z^2)+F(2+z)=2
\ee
with general solution
\be
F(z)=\frac{2\text{Arcosh}\left(\frac{z+2}{2}\right)}{\sqrt{4z+z^2}}
+\frac{c}{\sqrt{4z+z^2}}\label{F}
\ee
depending on an integration constant $c$. $I_f$ vanishes for $m\to\infty$
or $z\to0$, which entails $c=0$, producing (\ref{integral-1}).


Performing exactly the same steps as before we can calculate
the heat-kernel integral $I_{f-1}$ of (\ref{integral-2}) with
the function
\be
G(z)=\sum_{n=0}^{\infty}(-1)^nb_nz^n\,,\,\,b_n=\frac{n}{2(2n+3)}b_{n-1}.
\label{G}
\ee

\section{Third Order of CPTH}

To calculate the IR-divergent term
of order $E^3$ in CPTH we start from (see equation (2.28)
in \cite{bav87})\footnote{A d'Alembertian with index is supposed to act only
on functions with the same index, i.e. $\sq_1E_1:=\lim_{x_1\to x}\sq_{x_1}E(x_1)$.}
\small
\begin{multline}
\frac{d}{ds}\left\{\frac{1}{\Gamma(s)}\int_0^{\infty}\tau^{s-1}d\tau
\frac{\tau^3}{3\tau^1}\int_0^1d\ga\int_0^{\ga}d\gb\right.\\
\left.\cdot\int_{L_{\cal E}}\sqrt{g}d^2x
e^{-\tau\left[\gb(\ga-\gb)\sq_1+\gb(1-\ga)\sq_2+(\ga-\gb)(1-\ga)\sq_3+m^2\right]}
E_1E_2E_3\right\}\biggm|_{s=0}\\
=\frac{d}{ds}\left\{\int_{\ga}\int_{\gb}\int_{L_{\cal E}}
\left[\gb(\ga-\gb)\sq_1+\gb(1-\ga)\sq_2
+(\ga-\gb)(1-\ga)\sq_3+m^2\right]^{-s-2}\right.\\
\left.\cdot\frac{E_1E_2E_3}{3}\cdot s(s+1)\right\}\biggm|_{s=0}.
\label{third order}
\end{multline}
\normalsize
The $\gb$-integration is done first, using the formula 
\be
\int\frac{1}{X^n}dx=\frac{2ax+b}{(n-1)\D X^{n-1}}
+\frac{(2n-3)2a}{(n-1)\D}\int\frac{dx}{X^{n-1}},
\label{formula}
\ee
where we have $a=-\sq_1,b=\ga\sq_1+(1-\ga)[\sq_2-\sq_3],c=\ga(1-\ga)\sq_3+m^2$ and
$\D=-\ga^2\sq_1^2-(1-\ga)^2[\sq_2-\sq_3]^2-2\ga(1-\ga)\sq_1[\sq_2-\sq_3]-4m^2\sq_1$.
$m^2$ can be set to zero in $\D$, because it leaves the whole expression
regular. This can be checked by inserting $\ga={0,1}$. The $\gb$-integration yields
\begin{multline}
\int_0^{\ga}Y^{-s-2}d\gb=\frac{-2\sq_1\gb+\ga\sq_1+(1-\ga)[\sq_2-\sq_3]}
{(s+1)\D\cdot Y^{s+1}}\biggm|_0^{\ga}\\
-\frac{(2s+1)2\sq_1}{(s+1)\D}\int_0^{\ga}Y^{-s-1}d\gb,
\end{multline}
where $Y=\gb(\ga-\gb)\sq_1+\gb(1-\ga)\sq_2+(\ga-\gb)(1-\ga)\sq_3+m^2$.
Only the surface term retains the IR divergence while the remaining integral
will become regular after the $\ga$-integration. The upper limit
contribution $\gb=\ga$ is
\be
\frac{-\ga\sq_1+(1-\ga)[\sq_2-\sq_3]}{(s+1)\D[\ga(1-\ga)\sq_2+m^2]^{s+1}},
\ee
while the lower limit $\gb=0$ gives
\be
-\frac{\ga\sq_1+(1-\ga)[\sq_2-\sq_3]}{(s+1)\D[\ga(1-\ga)\sq_3+m^2]^{s+1}},
\ee
these contributions being symmetric under the exchange $2\leftrightarrow3$.
Next we perform the $\ga$-integration for the upper limit contribution, partially
integrating the IR divergent expression $[\ga(1-\ga)\sq_2+m^2]^{-s-1}$.
The remaining term, being differentiated, cannot lead to further
IR divergences because $\D^{-1}$ is finite on the whole $\ga$-interval.
For the partial $\ga$-integration we have $a=-\sq_2,b=\sq_2,c=m^2$,
$\D_{\ga}=-\sq_2^2+O(m^2)$ and $X=\ga(1-\ga)\sq_2+m^2$
\small
\begin{multline}
\int_0^1\frac{-\ga\sq_1+(1-\ga)[\sq_2-\sq_3]}{(s+1)\D}
[\ga(1-\ga)\sq_2+m^2]^{-s-1}d\ga\\
=\left\{\frac{-\ga\sq_1+(1-\ga)[\sq_2-\sq_3]}{(s+1)\D}
\left[\frac{-2\sq_2\ga+\sq_2}{-s\sq_2^2}X^{-s}+\frac{-(2s-1)2\sq_2}{-s\sq_2^2}
\int X^{-s}d\ga\right]\right\}_0^1\\
-\int_0^1\left[\frac{-2\sq_2\ga+\sq_2}{-s\sq_2^2}X^{-s}
+\frac{-(2s-1)2\sq_2}{-s\sq_2^2}\int X^{-s}d\ga\right]\\
\cdot\p_{\ga}\frac{-\ga\sq_1+(1-\ga)[\sq_2-\sq_3]}{(s+1)\D}d\ga
\label{alpha-int.}
\end{multline}
\normalsize
All terms contain a factor $[s(s+1)]^{-1}$ which is cancelled
by an identical one in (\ref{third order}). The differentiation for $s$ thus
only acts on factors $2s-1$ in the numerators, leaving such expressions
harmless, and on $X^{-s}$, thereby producing a logarithm of the argument.
If the latter is evaluated directly
on the boundary we arrive at a logarithmic divergence. The remaining integral
is regular. Hence, the logarithmic divergence only appears
in the first term of the second line in (\ref{alpha-int.}).
We first evaluate it at the boundary:
\begin{multline}
\left(\frac{-\ga\sq_1+(1-\ga)[\sq_2-\sq_3]}{(s+1)\D}\right)
\left(\frac{-2\sq_2\ga+\sq_2}{-s\sq_2^2}\right)[\ga(1-\ga)\sq_2+m^2]^{-s}\biggm|_0^1\\
=\left[\frac{1}{\sq_1\sq_2}+\frac{1}{(\sq_3-\sq_2)\sq_2}\right]\frac{(m^2)^{-s}}{s(s+1)}
\end{multline}
By the symmetry $2\leftrightarrow3$ the lower limit of the
$\gb$-integration yields an IR divergent term
\be
\left[\frac{1}{\sq_1\sq_3}+\frac{1}{(\sq_2-\sq_3)\sq_3}\right]
\frac{(m^2)^{-s}}{s(s+1)}
\ee
after the $\ga$-integration has been performed. Finally we put
these terms into (\ref{third order}) and obtain the unique IR divergent
contribution of the order $E^3$:
\begin{multline}
\frac{d}{ds}\left\{(m^2)^{-s}\int_{L_{\cal E}}\sqrt{g}d^2x
\left(\frac{1}{\sq_1\sq_2}
+\frac{1}{\sq_1\sq_3}+\frac{1}{\sq_2\sq_3}\right)
\frac{E_1E_2E_3}{3}\biggm|_{\{1,2,3\}=1}\right\}\\
\stackrel{s\to0}{\to}-\ln m^2\int_{L_{\cal E}}
E\frac{1}{\sq}E\frac{1}{\sq}E\sqrt{g}d^2x
\end{multline}

\section{Regularity of $\ln\sq$}

The regularity of the formal expression $\ln\sq$ in (\ref{I_f limit})
must be discussed for the combination $(\ln\sq/\sq)E$ which appears in the CPTH
to second and higher order. An explicit expression
can be found by inserting eigenfunctions $\vp_{\gl}(y)$
of the d'Alembertian. Since the action of $\ln\sq$ on the eigenfunctions
is not yet defined, we must go back to the function $F(z)$
(see eq. (\ref{I_f-exact}), Appendix A) which produces the
term $\propto(\ln\sq/\sq)E$ in the limit $m\to0$.
Expanding $z\cdot F(z)$ in a power series in $z=\sq/m^2$ its
action on the eigenfunctions is well-defined and yields (\ref{I_f E}).

Eigenfunctions and eigenvalues are defined as
$\sq\vp_{\gl}(y)=\tilde{\gl}^2\,\,\vp_{\gl}(y)=\gl^2/(4M^2)\,\,\vp_{\gl}(y)$
where $\gl$ is dimensionless (see below).
We have assumed that $E$ is time-independent as it is the case for SRG.
In a next step we would like to perform the limit $m^2\to0$ in the
integrand of (\ref{I_f E}) to isolate the IR divergence $\propto\ln m^2$.
Such a result, however, critically depends on the regularity of the $\gl$-integration,
in particular on the behavior of the eigenfunctions $\vp_{\gl}(y)$
in the limit $\gl\to0$.

\subsection{Eigenfunctions}

The eigenvalue equation of the radial d'Alembertian
\be
\sq_r\vp_{\gl}(r)=-\frac{d}{dr}\left[\Sss\frac{d}{dr}\right]
\vp_{\gl}(r)=\tilde{\gl}^2\cdot\vp_{\gl}(r)
\ee
with the dimensionless radius variable $y:=\frac{r}{2M}-1$
can be brought to dimensionless form
\be
-\frac{d}{dy}\left(\frac{y}{1+y}\,\frac{d}{dy}\right)\vp_{\gl}(y)
=\gl^2\cdot\vp_{\gl}(y)
\label{Laplace-Eigenequation}
\ee
with the dimensionless eigenvalue $\gl^2:=4M^2\tilde{\gl}^2$.
The differential equation (\ref{Laplace-Eigenequation}) possesses
two inessential singularities at $y=-1$ and $y=0$, but an essential
one at $y\to\infty$. Its solutions, therefore, do not belong
to Fuchs' class. At the horizon ($y=0$) eq. (\ref{Laplace-Eigenequation})
has two independent solutions, one of which is logarithmically divergent for $y\to0$.
They can be determined by standard methods as generalised
power series:
\ba
\vp_{\gl}^{(1)}&=&1-\gl^2 y+\frac{\gl^2(\gl^2-2)}{4}y^2-\frac{\gl^4(\gl^2-8)}{36}y^3+
O\left(y^4\right)\\
\vp_{\gl}^{(2)}&=&\vp_{\gl}^{(1)}\cdot\ln y+(2\gl^2+2)y-\frac{\gl^2(3\gl^2-2)}{4}y^2
+O\left(y^3\right)
\ea
Expanding the whole equation (rather than the solution)
for small values of $y$ the approximative equation $\vp'_{\gl}+\gl^2\vp_{\gl}=0$
has the unique solution
\be
\vp^h_{\gl}(y)\stackrel{y\to0}{\approx}c(\gl)\cdot e^{-\gl^2y}
=c(\gl)\cdot e^{-\gl^2(\frac{r}{2M}-1)},
\label{hor.solution}
\ee
which is regular at the horizon and approaches there the regular
one $\vp_{\gl}^{(1)}$ of the two solutions of the exact eigenvalue-equation
(\ref{Laplace-Eigenequation}). $c(\gl)$ is an unknown normalisation
factor. For large $y$ the solutions
\be
\vp^{\infty}_{\gl}(y)\stackrel{y\to\infty}{\approx}
a(\gl)\sin(\gl y)+b(\gl)\cos(\gl y)\label{asymp.solution}
\ee
correspond to the expected free wave at asymptotic distances.
In principle the normalisation factors $c(\gl)$, $a(\gl)$, and
$b(\gl)$ can be fixed by the orthonormality condition
\be
\int_0^{\infty}\vp_{\gl'}\vp_{\gl}dy=
\frac{1}{\D\gl^2}\frac{y}{1+y}
\left[\vp_{\gl'}\left(\vp'_{\gl}\right)
-\left(\vp'_{\gl'}\right)\vp_{\gl}\right]
\biggm|_0^{\infty}:=\delta(\D\gl)\label{orthogonality},
\ee
following from (\ref{Laplace-Eigenequation}), where $\D\gl=\gl'-\gl$
and $\D\gl^2=(\gl')^2-\gl^2$.
Because of the pre-factor $y/(1+y)$ only the upper boundary contributes
where $\vp_{\gl}^{(1)}$, the regular solution, behaves as (\ref{asymp.solution}).
It is convenient to compare (\ref{orthogonality})
to the orthogonality condition of the exactly known eigenfunctions
$\vp^0_{\gl}=a(\gl)\sin(\gl y)+b(\gl)\cos(\gl y)$
of the flat radial d'Alembertian $\sq_0=-\frac{d^2}{dy^2}$:
\begin{multline}
\int_0^{\infty}\vp^0_{\gl'}\vp^0_{\gl}dy=
\frac{1}{\D\gl^2}\left[\vp^0_{\gl'}\left(\frac{d}{dy}\vp^0_{\gl}\right)
-\left(\frac{d}{dy}\vp^0_{\gl'}\right)\vp^0_{\gl}\right]\biggm|_0^{\infty}\\
=\frac{\pi}{2}[a^2(\gl)+b^2(\gl)]\delta(\D\gl)
+\frac{\gl'a(\gl')b(\gl)-\gl a(\gl)b(\gl')}{\D\gl^2}
\label{orthogonality2}
\end{multline}
The appearance of the last term in (\ref{orthogonality2}) expresses
the fact that the flat eigenfunctions
$\vp^0_{\gl}$ are \emph{not} orthogonal on the half-line.
More precisely, the $\sin(\gl y)$-modes are not orthogonal to the
$\cos(\gl y)$-modes\footnote{For instance, the term $a(\gl')b(\gl)
\int_0^{\infty}\sin(\gl'y)\sin(\gl y)dy$ has been calculated by differentiating the identity
$\int_0^{\infty}\sin y\cos(a y)/y dy=\pi/4[1-\sgn(a-1)]$ for $a$.
For the term $a(\gl')b(\gl)\int_0^{\infty}\sin(\gl'y)\cos(\gl y)dy$, spoiling
the orthogonality, we differentiate $\int_0^{\infty}[\cos(ay)-\cos(by)]/ydy=\ln b/a$
for $a$ and set $a=\gl-\gl',b=\gl+\gl'$.}.

The upper-limit contributions of (\ref{orthogonality})
and (\ref{orthogonality2}) are identical. The lower limit $y=0$
in (\ref{asymp.solution}) vanishes for $\vp_{\gl}$. Thus we can write
\begin{multline}
\int_0^{\infty}\vp_{\gl'}\vp_{\gl}dy=\int_0^{\infty}\vp^0_{\gl'}\vp^0_{\gl}dy
+\frac{1}{\D\gl^2}\left[\vp^0_{\gl'}\left(\p_y\vp^0_{\gl}\right)
-\left(\p_y\vp^0_{\gl'}\right)\vp^0_{\gl}\right]\biggm|^0\\
=\frac{\pi}{2}[a^2(\gl)+b^2(\gl)]\delta(\D\gl).\label{normalisation}
\end{multline}
The lower-limit contribution (which can be calculated directly
inserting the flat eigenfunctions $\vp^0_{\gl}$) just happens to
cancel the inconvenient last term in (\ref{orthogonality2}).
Therefore, the exact eigenfunctions $\vp_{\gl}$ of the curved d'Alembertian
are orthogonal on the half-line\footnote{A general Sturm-Liouville
operator of the type $-\frac{d}{dy}(g(y)\frac{d}{dy})$ shares this property,
if the function $g(y)$ satisfies $g(\infty)=1$ and $g(0)=0$
as in our case (\ref{Laplace-Eigenequation}).} $y\in[0,\infty[$.
According to (\ref{normalisation}) the normalisation is fixed by a choice
\be
a(\gl)=\sqrt{\frac{2}{\pi}}\cdot\hat{a}\,\,,
\,\,b(\gl)=\sqrt{\frac{2}{\pi}}\cdot\hat{b}
\ee
for the factors of the large $y$ solutions (\ref{asymp.solution}),
where $\hat{a}^2+\hat{b}^2=1$.
The normalisation $c(\gl)$ of the unique exponential solution (\ref{hor.solution})
at small values of $y$, however, is still unknown.
As the differential equation (\ref{Laplace-Eigenequation}) does not belong
to Fuchs' class, the relation between solutions at different singularities
is not known. Instead we try to find an answer by an
approximate patching of the solutions at some intermediate value
$\bar{y}$. Considering (\ref{Laplace-Eigenequation}) it is evident that the
solutions $\vp^h_{\gl},\vp^{\infty}_{\gl}$ are accurate for $y\ll1\,,\,y\gg1$,
respectively. Therefore, the choice $\bar{y}=1$ seems appropriate
for the patching. We assume that at this point the regular solution
(\ref{hor.solution}) and its derivative with
$c(\gl)=\sqrt{2/\pi}\,\hat{c}$ shall approach a linear
combination of the two large $y$ solutions:
\ba
\hat{c}\cdot e^{-\gl^2}&\approx&
\left[\hat{a}\sin(\gl)+\hat{b}\cos(\gl)\right]\\
-\gl^2\,\hat{c}\cdot e^{-\gl^2}&\approx&\gl
\left[\hat{a}\cos(\gl)-\hat{b}\sin(\gl)\right]
\ea
The normalisations are fixed by the limit $\gl\to0$,
leading to the equations:
\ba
\hspace{-0.6cm}\hat{c}\left[1-\gl^2+O\left(\gl^4\right)\right]&\approx&
\hat{a}\left[\gl-\frac{\gl^3}{6}+O\left(\gl^5\right)\right]
+\hat{b}\left[1-\frac{\gl^2}{2}+O\left(\gl^4\right)\right]\\
\hspace{-0.6cm}\hat{c}\left[1-\gl^2+O\left(\gl^4\right)\right]&\approx&
-\hat{a}\left[\frac{1}{\gl}-\frac{\gl}{2}+O\left(\gl^3\right)\right]
+\hat{b}\left[1-\frac{\gl^2}{6}+O\left(\gl^4\right)\right]
\ea
Keeping the lowest orders of $\gl$ only, the best approximative solution
to these equations is given by $\hat{a}=0,\hat{c}=\hat{b}=1$.
Therefore, the (regular) normalised eigenfunctions
near the horizon and asymptotically should behave like
\be
\vp_{\gl}(y)\propto\sqrt{\frac{2}{\pi}}\cdot\left\{
\begin{array}{ll}
e^{-\gl^2 y} & ,\,\,y\to0 \\
\cos(\gl y)
& ,\,\,y\to\infty.
\end{array}\right.
\label{approx-solutions}
\ee

\subsection{Regularity}

In $I_f E$ of (\ref{I_f E}) the behavior of the eigenfunctions (\ref{approx-solutions})
and of $F(\frac{\gl^2}{4M^2m^2})$ especially at small valus of
$\gl$ seems to be sufficiently nice, so that the limit $m^2\to0$ can be performed
and the divergence $\propto\ln m^2$ can be isolated:
\be
I_f E\stackrel{m\to0}{\to}\frac{8M^2}{\sq}
\int_0^{\infty}d\gl\ln\left(\frac{\gl^2}{4M^2m^2}\right)
\,\,\vp_{\gl}(y)\int_0^{\infty}\vp_{\gl}(y')\cdot E(y')dy'
\ee
A necessary condition for that is the regularity of\footnote{For the moment
we neglect the action of $\sq^{-1}$ in $I_f$. Pulling it into
the $\gl$-integral would produce the typical IR problem of the
two-dimensional Green function whose resolution shall not be discussed
here (cf. the last paragraph of this Appendix).}
\be
I^{reg}_f E\propto
\int_0^{\infty}d\gl\ln\gl
\,\,\vp_{\gl}(y)\int_0^{\infty}\vp_{\gl}(y')\cdot E(y')dy'.
\label{I_f->E}
\ee
The integral over $y'$ can be split into the domains of the
approximative solutions (\ref{approx-solutions}):
\be
E_{\gl}:=\int_0^{\infty}\vp_{\gl}(y')\cdot E(y')dy'
\approx\int_0^1 e^{-\gl^2 y'}E(y')dy'
+\int_1^{\infty}\cos(\gl y')E(y')dy'
\ee
In SRG, where $E=\frac{2M}{r^3}\propto\frac{1}{(y+1)^3}$, we can
approximate these integrals by the functions
\ba
\int_0^1\frac{e^{-\gl^2 y'}}{(y'+1)^3}dy'&\approx&\frac{1}{\gl^2+\frac{8}{3}}
\label{y'=0}\\
\int_1^{\infty}\frac{\cos(\gl y')}{(y'+1)^3}dy'&\approx&
\left(2e^{-\gl}-1\right)\frac{\sin(\gl)}{8 \gl}
\label{y'=infty}
\ea
which correctly reproduce the behavior of the integrals
for small and large values of $\gl$. For intermediate values
the deviations are also small as we have checked graphically.
The function
\be
E_{\gl}\approx\frac{1}{\gl^2+\frac{8}{3}}+\left(2e^{-\gl}-1\right)\frac{\sin(\gl)}{8 \gl}
\label{E_l}
\ee
is regular for all values of $\gl$, taking the value
$1/2$ at $\gl=0$ and behaving like $\sin \gl/\gl$ at infinity.
Inserting (\ref{E_l}) into (\ref{I_f->E}) the behavior
of $I^{reg}_f E$ at either the horizon ($y\to0$) or asymptotically
($y\to\infty$) can be investigated.

At the horizon only the part $\sin(\gl)/\gl$ of $E_{\gl}$ may cause problems
in (\ref{I_f->E}) for $\gl\to\infty$. However,
$\int_0^{\infty}d\gl(\ln \gl\sin \gl)/\gl$ is
finite, thus establishing regularity of (\ref{I_f->E}) for $y\to0$.

To show the asymptotic behavior of $I^{reg}_f E$ is more tedious.
First we consider the integral $\int_0^{\infty}\ln \gl\cos(\gl y)/(\gl^2+8/3)d\gl$
from (\ref{y'=0}) in (\ref{I_f->E}) in the limit $y\to\infty$.
We split it at the point $\gl=1$. The first contribution is
\begin{multline}
\int_0^1\frac{\ln \gl\cos(\gl y)}{\gl^2+\frac{8}{3}}d\gl
=\frac{1}{y}\int_0^y\frac{\ln\left(\frac{s}{y}\right)\cos s}{\frac{s}{y}^2+\frac{8}{3}}ds
<\frac{3}{8 y}\int_0^y\ln\left(\frac{s}{y}\right)\cos sds\\
=\frac{3}{8 y}\left[\ln\left(\frac{s}{y}\right)\sin s\biggm|_0^y
-\int_0^y\frac{\sin s}{s}ds\right]=\frac{3\pi}{16\,\,y}+O(y^{-2}),
\end{multline}
where a substitution $s=\gl y$ has been performed. The second
contribution by 
\begin{multline}
\biggm|\int_1^{\infty}\frac{\ln \gl\cos(\gl y)}{\gl^2+\frac{8}{3}}d\gl\biggm|\\
=\frac{1}{y}\biggm|\left\{\frac{\ln \gl\sin(\gl y)}{\gl^2+\frac{8}{3}}\biggm|_1^{\infty}
-\int_1^{\infty}\sin(\gl y)\left[\frac{1}{\gl\left(\gl^2+\frac{8}{3}\right)}
-\frac{2\gl\ln \gl}{\left(\gl^2+\frac{8}{3}\right)^2}\right]d\gl\right\}\biggm|\\
<\frac{1}{y}\left\{\biggm|\int_1^{\infty}\frac{1}{\gl\left(\gl^2+\frac{8}{3}\right)}d\gl\biggm|
+\biggm|\int_0^1\frac{2\gl\ln \gl}{\left(\gl^2+\frac{8}{3}\right)^2}d\gl\biggm|\right\}
=\frac{3\ln(11/3)}{8\,\,y}
\end{multline}
is also at most of order $y^{-1}$. Next we consider the integral\linebreak
$\int_0^{\infty}\ln \gl\cos(\gl y)\left(2e^{-\gl}-1\right)\sin(\gl)/(8 \gl)d\gl$.
To obtain an upper bound it is sufficient to set
$\left(2e^{-\gl}-1\right)\to1$:
\begin{multline}
\int_0^{\infty}\frac{\ln \gl\cos(\gl y)\sin \gl}{\gl}=\frac{1}{2}\int_0^{\infty}
\frac{\ln \gl\left\{\sin[\gl(y+1)]-\sin[\gl(y-1)]\right\}}{\gl}d\gl\\
=\frac{1}{2}\int_0^{\infty}\frac{\left\{\ln\left(\frac{s}{y+1}\right)
\sin s-\ln\left(\frac{s}{y-1}\right)\sin s\right\}}{s}ds\\
=\frac{1}{2}\ln\left(\frac{y-1}{y+1}\right)\int_0^{\infty}\frac{\sin s}{s}ds
\stackrel{y\to\infty}{\to}0-\frac{\pi}{2\,\,y}+O(y^{-2})
\end{multline}
From the first to the second line we have performed a substution
$s=\gl(y\pm1)$, respectively.

The present analysis can only be seen as some basic argument
for the regularity of $\ln\sq E$. In the effective action
one has to deal with terms $E\sq^{-1}\ln\sq E$, where the $\sq^{-1}$
can be made acting to the left. In the case of expectation values
resulting from variations for $E$, however, one is forced to evaluate
$I_fE$ as it stands. In this case an explicit analysis of the
exact Green function (in terms of some expansion) seems inevitable.

\end{appendix}

\end{document}